\documentclass[article]{IEEEtran}

\usepackage{graphicx}
\usepackage{amsmath,amsfonts,amssymb,amsthm}
\usepackage{cite}

\title{DyCaPPON: Dynamic Circuit and Packet Passive Optical
   Network (Extended Version)\thanks{Technical Report, School of
Electrical, Computer, and Energy Eng., Arizona State Univ., April 2014.
This extended technical report accompanies~\cite{WeAMR14}.}}

\author{Xing~Wei, Frank Aurzada, Michael~P.~McGarry, and Martin~Reisslein
\thanks{X.~Wei and M.~Reisslein are with
the School of Electrical, Computer, and Energy Engineering,
Arizona State University, Tempe, AZ 85287-5706,
Email: \{Xing.Wei, reisslein\}@asu.edu,
Phone: (480) 965-8593, Fax: (480) 965-8325.}
\thanks{F.~Aurzada is with the Mathematics Faculty at the
Technical University Darmstadt, Schlossgartenstr.~7,
64289 Darmstadt, Germany, Email: aurzada@mathematik.tu-darmstadt.de,
Phone: + 49 6151 16-3183, Fax: + 49 6151 16-6822}
\thanks{M.\ McGarry is with
the Dept. of Electrical and Computer Eng., University of Texas at El Paso,
El Paso, TX, Email: mpmcgarry@utep.edu,
Phone: (915) 747-6955, Fax: (915) 747-7871.}
 }

\def\E{\mathbb{E}}

\begin{document}

\maketitle

\begin{abstract}
Dynamic circuits are well suited for applications that require
predictable service with a
constant bit rate for a prescribed period of time, such as
cloud computing and e-science applications.
Past research on upstream transmission in
passive optical networks (PONs) has mainly
considered packet-switched traffic and has focused on optimizing
packet-level performance metrics, such as reducing mean delay.
This study proposes and evaluates a dynamic circuit and packet
PON (DyCaPPON) that provides dynamic circuits along with packet-switched
service.
DyCaPPON provides $(i)$ flexible packet-switched service through
dynamic bandwidth allocation in periodic polling cycles,
and $(ii)$ consistent circuit service by allocating each active
circuit a fixed-duration upstream transmission window during
each fixed-duration polling cycle.
We analyze circuit-level performance metrics, including
the blocking probability of dynamic circuit requests
in DyCaPPON through
a stochastic knapsack-based analysis.
Through this analysis we also determine the bandwidth occupied
by admitted circuits. The remaining bandwidth is available for packet
traffic and we conduct an approximate analysis of
the resulting mean delay of packet traffic.
Through extensive numerical evaluations and verifying simulations
we demonstrate the circuit blocking and packet delay trade-offs in
DyCaPPON.
\end{abstract}

\begin{keywords}
Dynamic circuit switching; Ethernet Passive Optical Network;
Grant scheduling; Grant sizing; Packet delay; Stochastic knapsack.
\end{keywords}

\section{Introduction}
\label{sec:intro}
Optical networks have traditionally employed
three main switching paradigms, namely
circuit switching, burst switching, and packet
switching, which have extensively studied respective benefits and
limitations~\cite{PerformCS,OCSvsOBS01,OVSvsOBSpkt,NEHONET}.
In order to achieve the predictable network service of circuit switching
while enjoying some of the flexibilities of burst and packet
switching, \textit{dynamic circuit switching} has been
introduced~\cite{VeKC10}.
Dynamic circuit switching can be traced back to research
toward differentiated levels of blocking rates of calls~\cite{DCS1989}.
Today, a plethora of network applications ranging from
the migration of data and computing work loads to cloud storage and
computing~\cite{SBCD1009}
as well as high-bit rate e-science applications, e.g., for
remote scientific collaborations, to big data applications of
governments, private organizations, and households
are well supported by dynamic circuit switching~\cite{VeKC10}.
Moreover, gaming applications benefit from predictable
low-delay service~\cite{BrF10,FiGR02,MaH10,ScER02} provided by circuits,
as do emerging virtual reality
applications~\cite{KuB13,PaDR12,VaZK10}.
Also, circuits can aid in the timely transmission of
data from continuous media applications, such as live or
streaming video.
Video traffic is often highly variable and
may require smoothing before transmission over
a circuit~\cite{ghazi2012vmp,oh2008cont,QiKo11,ReLR02,ReT99,ShSZ11,AuRe09}
or require a combination of circuit transport for a constant
base bit stream and packet switched transport for the traffic burst
exceeding the base bit stream rate.
Both commercial and
research/education network providers have recently started
to offer optical dynamic circuit switching services~\cite{DOCS00,DCN}.

While dynamic circuit switching
has received growing research attention in core and metro
networks~\cite{DCN,Charb12,HybridN2010,Li08,MGJT0511,ReqprovDOCS,Sko12,VanBreu05,CircuitSONET},
mechanisms for supporting dynamic circuit switching in
passive optical networks (PONs), which are a
promising technology for network
access~\cite{Mahloo13,McR12,McRAS10,Siva13,ToKK12,Zan2013},
are largely an open research area.
As reviewed in Section~\ref{lit:sec}, PON research
on the upstream transmission direction from the distributed
Optical Network Units (ONUs) to the central Optical Line Terminal (OLT) has
mainly focused on mechanisms supporting packet-switched
transport~\cite{AuSH08,AuSRGM11,ZhMo09}.
While some of these packet-switched transport mechanisms
support quality of service akin to circuits through
service differentiation mechanisms, to the best of our knowledge
there has been no prior study of circuit-level performance
in PONs, e.g., the blocking probability of circuit requests
for a given circuit request rate and circuit holding time.

In this article, we present the first circuit-level performance
study of a PON with polling-based medium access control.
We make three main original contributions towards the concept of efficiently
supporting both \textbf{Dy}namic \textbf{C}ircuit \textbf{a}nd \textbf{P}acket
traffic in the upstream
direction on a \textbf{PON}, which we refer to as \textbf{DyCaPPON}:
\begin{itemize}
\item We propose a novel DyCaPPON polling cycle structure that exploits
the dynamic circuit transmissions to mask the round-trip propagation delay
for dynamic bandwidth allocation to packet traffic.
\item We develop a stochastic knapsack-based model of DyCaPPON
to evaluate the circuit-level performance, including
the blocking probabilities for different classes of
circuit requests.
\item We analyze the bandwidth sharing between circuit and packet traffic
in DyCaPPON
and evaluate packet-level performance, such as mean packet delay,
as a function of the circuit traffic.
\end{itemize}

This article is organized as follows.
We first review related work in Section~\ref{lit:sec}.
In Section~\ref{sec:model}, we describe the considered
access network structure and define both the circuit and packet traffic models
as well as the corresponding circuit- and packet-level performance metrics.
In Section~\ref{dycappon:sec}, we introduce the DyCaPPON polling
cycle structure and outline the steps for admission control of
dynamic circuit requests and dynamic bandwidth allocation to packet traffic.
In Section~\ref{sec:analysis} we analyze the performance metrics
relating to the dynamic circuit traffic, namely the blocking
probabilities for the different circuit classes. We also
analyze the bandwidth portion of a cycle consumed by active circuits,
which in turn determines the bandwidth portion available for packet traffic,
and analyze the resulting mean delay for packet traffic.
In Section~\ref{eval:sec} we
validate numerical results from our analysis with simulations and present
illustrative circuit- and packet-level performance results for DyCaPPON.
We summarize our conclusions in Section~\ref{sec:conclusion}
and outline future research directions towards the DyCaPPON concept.

\section{Related Work}
\label{lit:sec}
The existing research on upstream transmission in
passive optical access networks has mainly focused on
packet traffic and related packet-level performance metrics.
A number of studies has primarily focused on
differentiating the packet-level QoS for different classes
of packet traffic,
e.g.,~\cite{AnLA04,AsYDA03,DiDLC11,GhSA04,LuAn05,RaM09,
ShHEAA04,ShBGAM05,VaGh11}.
In contrast to these studies, we consider only best effort service
for the packet traffic in this article.
In future work, mechanisms for differentiation of packet-level QoS
could be integrated into the packet partition
(see Section~\ref{dycappon:sec}) of the DyCaPPON polling cycle.

The needs of applications for transmission
with predictable quality of service has led to various
enhancements of packet-switched transport for providing quality of
service (QoS).
A few studies, e.g.,~\cite{BeBM09,Ho06,MZC0303,Qin13,ZhAY03,ZhP04},
have specifically focused on providing deterministic QoS, i.e.,
absolute guarantees for packet-level performance metrics,
such as packet delay or jitter.
Several studies have had a focus on the efficient integration of
deterministic QoS mechanisms with one or several
lower-priority packet traffic classes in polling-based
PONs, e.g.,\cite{AHKWK0703,BeIB11,DhAMS07,HwLLL12,LiLC11,Merayo2010,NgGB11}.
The resulting packet scheduling problems have received
particular attention~\cite{De12,PeFA09,YiP10}.
Generally, these prior studies have found that fixed-duration
polling cycles are well suited for supporting consistent
QoS service.
Similar to prior studies, we employ fixed-duration polling
cycles in DyCaPPON, specifically on a PON with a single-wavelength
upstream channel.

The prior studies commonly considered traffic flows characterized
through leaky-bucket parameters that bound the long-term average
bit rate as well as the size of sudden traffic bursts.
Most of these studies include admission control, i.e., admit
a new traffic flow only when the packet-level performance
guarantees can still be met with the new traffic flow
added to the existing flows.
However, the circuit-level performance, i.e., the probability of
blocking (i.e., denial of admission) of a new request has not
been considered.
In contrast, the circuits in
DyCaPPON provide absolute QoS to constant bit rate
traffic flows without bursts and we analyze the probability
of new traffic flows (circuits) being admitted or blocked.
This flow (circuit) level performance is important
for network dimensioning and providing QoS at the level of
traffic flows.

For completeness, we briefly note that a PON
architecture that can provide circuits to ONUs through
orthogonal frequency division multiplexing techniques on the
physical layer has been proposed in \cite{OFDMPON}.
Our study, in contrast, focuses on efficient medium access control
techniques for supporting circuit traffic.
A QoS approach based on burst switching in a PON has been proposed
in~\cite{SeBP05}.
To the best of our knowledge, circuit level performance
in PONs has so far only been examined in~\cite{VaML12} for
the specific context of optical code division
multiplexing~\cite{KwYZ96}.

We also note for completeness that large file transmissions
in optical networks have been examined in~\cite{DSAlgfle},
where scheduling of large
data file transfers on the optical grid network is studied,
in~\cite{LgfleMulp}, where
parallel transfer over multiple network paths are examined, and
in~\cite{EIBT}, where files are transmitted in
a burst mode, i.e., sequentially.

Sharing of a general time-division multiplexing (TDM) link by
circuit and packet traffic has been analyzed in several
studies,
e.g.~\cite{bolla97,gaver82,ghani1994decomp,li1985perf,mag82,mankus92,wein80}.
These queueing theoretic analyses typically employed detailed Markov
models and become computationally quite demanding for high-speed
links. Also, these complex existing models considered a given node
with local control of all link transmissions. In contrast, we
develop a simple performance model for the distributed transmissions
of the ONUs that are coordinated through polling-based medium access
control in DyCaPPON. Our DyCaPPON model is accurate for the circuits
and approximate for the packet service. More specifically, we model
the dynamics of the circuit traffic, which is given priority over
packet traffic up to an aggregate circuit bandwidth of $C_c$ in
DyCaPPON, with accurate stochastic knapsack modeling techniques in
Section~\ref{percir:sec}. In Section~\ref{pkt_perf:sec}, we present
an approximate delay model for the packet traffic, which in DyCaPPON
can consume the bandwidth left unused by circuit traffic.

\section{System Model}
\label{sec:model}

\begin{table}
\caption{Main model notations}
\label{not:tab}
\begin{tabular}{|l|l|}  \hline
\multicolumn{2}{|c|}{Network architecture}\\
$C$  & Transmission rate [bit/s] of upstream channel  \\
$C_c$ & Transm. rate limit for circuit service, $C_c \leq C$ \\
$J$  & Number of ONUs \\
$\tau$  & One-way propagation delay [s] \\
\hline
\multicolumn{2}{|c|}{Traffic model}\\
$\mathbf{b} = (b_1,\ldots, b_K)$ & Bit rates [bit/s] for
         circuit classes $k = 1, 2, \ldots, K$\\
$\lambda_c$ & Aggregate circuit requests arrival rate [circuits/s]\\
$p_k$ & Prob. that a request is for circuit type $k$ \\
$\bar{b} = \sum_{k = 1}^K p_k b_k$ & Mean circuit bit rate [bit/s] of offered
     circuit traf. \\
$1/\mu$ & Mean circuit holding time [s/circuit] \\
$\chi = \frac{\lambda_c \bar{b}}{\mu C}$ & Offered circuit traffic
            intensity (load) \\
$\bar{P}$, $\sigma_p^2$ & Mean [bit] and variance of packet size \\
$\pi = \frac{\lambda_p \bar{P}}{C}$ & Packet traffic intensity (load);
 $\lambda_p$ is agg.\ packet  \\
   & \ \ \ \ \ generation rate [packets/s] at all $J$ ONUs\\
\hline
\multicolumn{2}{|c|}{Polling protocol}\\
$\Gamma$ & Total cycle duration [s], constant \\
$\Xi$ & Cycle duration (rand.\ var.) occupied by circuit traf. \\
$\omega$ & Mean per-cycle overhead time [s] for upstream \\
   & transmissions (report transm.\ times, guard times) \\
\hline
\multicolumn{2}{|c|}{Stochastic knapsack model for circuits}\\
$\mathbf{n} = (n_1, \ldots, n_K)$  & State vector of numbers of circuits
       of class $k$\\
$\beta = \mathbf{n} \cdot \mathbf{b}$  & Aggregate bandwidth of active
 circuits\\
$q(\beta)$  & Equilibrium probability for active circuits having \\
        & \ \ \ \ \ aggregate
    bandwidth $\beta$ \\ \hline
\multicolumn{2}{|c|}{Performance metrics}\\
$B_k$ & Blocking probability for circuit class $k$ \\
$D$ & Mean packet delay [s]\\
\hline
\end{tabular}
\end{table}
\subsection{Network structure}
We consider a PON with $J$ ONUs attached to the OLT with a single
downstream wavelength channel and a single upstream wavelength
channel~\cite{ZhMo09,MRM1008}.
We denote $C$ for the transmission bit rate (bandwidth) of a channel [bits/s].
We denote $\tau$ [s] for the one-way propagation delay between
the OLT and the equidistant ONUs.
We denote $\Gamma$ [s] for the fixed duration of a polling cycle.
The model notations are summarized in Table~\ref{not:tab}.

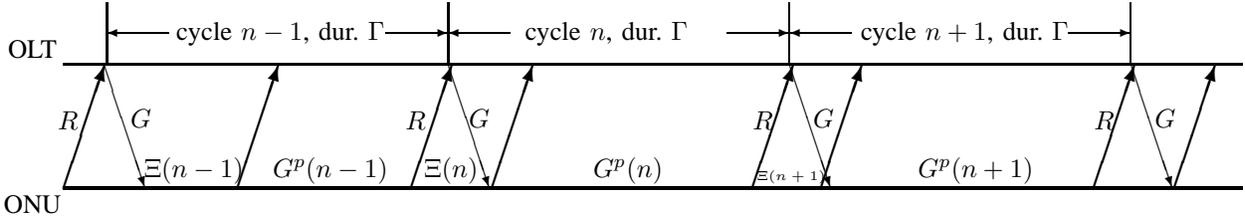
\begin{figure*}[t]
\begin{center}
\setlength{\unitlength}{0.825mm}
\begin{picture}(160,32)
\thicklines
\put(-10,0){\line(1,0){190}}
\put(-10,20){\line(1,0){190}}
\put(-15,21){\makebox(0,0)[b]{OLT}}
\put(-15,-1){\makebox(0,0)[t]{ONU}}

\thinlines
\put(-5,20){ \vector(1,-3){6.6}}
\thicklines
\put(-10,0){\vector(1,3){6.6}}
\put(18,0){\vector(1,3){6.6}}

\thinlines
\put(52,20){\vector(1,-3){6.6}}
\thicklines
\put(46,0){\vector(1,3){6.6}}
\put(59,0){\vector(1,3){6.6}}

\thinlines
\put(107,20){\vector(1,-3){6.6}}
\thicklines
\put(101,0){\vector(1,3){6.6}}
\put(112,0){\vector(1,3){6.6}}

\thinlines
\put(162,20){\vector(1,-3){6.6}}
\thicklines
\put(156,0){\vector(1,3){6.6}}
\put(169,0){\vector(1,3){6.6}}

\put(11,0.75){\makebox(0,0)[b]{$\Xi(n-1)$}}
\put(2.5,9.75){\makebox(0,0)[b]{$G$}}

\put(33,0.75){\makebox(0,0)[b]{$G^p(n-1)$}}

\put(52.5,0.75){\makebox(0,0)[b]{$\Xi(n)$}}
\put(57,9.75){\makebox(0,0)[b]{$G$}}

\put(81,0.75){\makebox(0,0)[b]{$G^p(n)$}}

\put(107.25,0.75){\makebox(0,0)[b]{\tiny{$\Xi(n+1)$}}}
\put(112.5,9.75){\makebox(0,0)[b]{$G$}}

\put(137,0.75){\makebox(0,0)[b]{$G^p(n+1)$}}

\put(167.5,9.75){\makebox(0,0)[b]{$G$}}

\thinlines
\put(-3,20){\line(0,1){10}}
\put(52,20){\line(0,1){10}}
\put(107,20){\line(0,1){10}}
\put(162,20){\line(0,1){10}}

\put(-10,11){\makebox(0,0)[]{ $R$}}
\put(46,11){\makebox(0,0)[]{ $R$}}
\put(102,11){\makebox(0,0)[]{ $R$}}
\put(157,11){\makebox(0,0)[]{ $R$}}

\put(7,25){\vector(-1,0){10}}
\put(42,25){\vector(1,0){10}}
\put(25,25){\makebox(0,0)[]{cycle $n-1$, dur.~$\Gamma$}}

\put(62,25){\vector(-1,0){10}}
\put(97,25){\vector(1,0){10}}
\put(77.5,25){\makebox(0,0)[]{cycle $n$, dur.~$\Gamma$}}

\put(117,25){\vector(-1,0){10}}
\put(152,25){\vector(1,0){10}}
\put(135.25,25){\makebox(0,0)[]{cycle $n+1$, dur.~$\Gamma$}}

\end{picture}
\end{center}
\caption{An upstream cycle $n$ has fixed duration $\Gamma$
  and has a circuit partition of duration $\Xi(n)$
  (that depends on the bandwidth demands of the accepted circuits)
  while a packet partition occupies the remaining cycle duration
   $\Gamma - \Xi(n)$.
The exact duration $G_p(n)$ of the packet partition in cycle $n$ is
evaluated in Eqn.~(\ref{Gp:eqn}).
 Each ONU sends a report during each packet partition.
Packet traffic reported in cycle $n-1$ is served in the packet partition
of cycle $n$ (if there is no backlog).
A circuit requested in cycle $n-1$ starts in the circuit partition of
cycle $n+1$.
The $2 \tau$ round-trip propagation delay between the last ONU report (R)
of a cycle $n-1$
and the first packet transmission following the grant (G)
of the next cycle $n$ is masked
by the circuit partition, provided $\Xi(n) > 2 \tau$. }
\label{fig:cycle}
\end{figure*}
\subsection{Traffic Models} \label{sec:trafficmod}
For circuit traffic, we consider $K$ classes of circuits
with bandwidths $\mathbf{b} = (b_1,\ b_2, \ldots, b_K)$.
We denote $\lambda_c$ [requests/s] for the aggregate Poisson process
arrival rate of circuit requests.
A given circuit request is for a circuit of class $k,\ k = 1, 2, \ldots, K$,
with probability $p_k$.
We denote the mean circuit bit rate of the offered circuit traffic by
$\bar{b} = \sum_{k=1}^K p_k b_k$.
We model the circuit holding time (duration) as an exponential random
variable with mean $1/\mu$.
We denote the resulting offered circuit traffic
intensity (load) by $\chi = \lambda_c \bar{b} / (\mu C)$.

For packet traffic, we denote $\bar{P}$ and $\sigma_p^2$ for the mean
and the variance of the packet size [in bit], respectively.
We denote $\lambda_p$ for the aggregate Poisson process
arrival rate [packets/s]
of packet traffic across the $J$ ONUs and denote
$\pi := \bar{P} \lambda_p/C $ for
the packet traffic intensity (load).

Throughout, we define the packet sizes and circuit bit rates to include the
per-packet overheads, such as the preamble for Ethernet frames
and the interpacket gap, as well as the packet overheads when
packetizing circuit traffic for transmission.

\subsection{Performance Metrics}
For circuit traffic, we consider the blocking probability
$B_k,\ k = 1, 2, \ldots, K$, i.e., the probability that a request
for a class $k$ circuit is blocked, i.e., cannot be accommodated within
the transmission rate limit for circuit service $C_c$.
We define the average circuit blocking probability as
$\bar{B} = \sum_{k = 1}^K p_k B_k$.
For packet traffic, we consider the mean packet
delay $D$ defined as
the time period from the instant of packet arrival at the ONU
to the instant of complete delivery of the packet to the OLT.

\section{DyCaPPON Upstream Bandwidth Management}
\label{dycappon:sec}
\subsection{Overview of Cycle and Polling Structure}
In order to provide circuit traffic with consistent upstream transmission
service with a fixed circuit bandwidth, DyCaPPON employs a polling cycle
with a fixed duration $\Gamma$ [s].
An active circuit with bandwidth $b$ is allocated an upstream
transmission window of duration $b \Gamma/C$ in every cycle.
Thus, by transmitting at the full upstream channel bit rate $C$
for duration $b \Gamma /C$ once per cycle of duration $\Gamma$,
the circuit experiences a transmission bit rate
(averaged over the cycle duration) of $b$.
We let $\Xi(n)$ denote the aggregate of the upstream
transmission windows of all active circuits in the PON in cycle $n$,
and refer to $\Xi(n)$ as the circuit partition duration.
We refer to the remaining duration $\Gamma - \Xi(n)$ as the
packet partition of cycle $n$.

As illustrated in Fig.~\ref{fig:cycle}, a given cycle $n$ consists
of the circuit partition followed by the packet partition. During
the packet partition of each cycle, each ONU sends a report message
to the OLT. The report message signals new circuit requests as well
as the occupancy level (queue depth) of the packet service queue in
the ONU to the OLT. The signaling information for the circuit
requests, i.e., requested circuit bandwidth and duration, can be
carried in the Report message of the MPCP protocol in EPONs with
similar modifications as used for signaling information for
operation on multiple wavelength channels~\cite{McGR06}.

Specifically, for signaling dynamic circuit requests,
an ONU report in the packet partition of cycle $n-1$
carries circuit requests generated since the ONU's preceding
report in cycle $n-2$. The report reaches the
OLT by the end of cycle $n-1$ and the OLT executes
circuit admission control as described in Section~\ref{ac:sec}.
The ONU is informed about the outcome of the admission control
(circuit is admitted or blocked)
in the gate message that is transmitted on the downstream wavelength
channel at the beginning of cycle $n$.
In the DyCaPPON design, the gate message propagates downstream while
the upstream circuit transmissions of cycle $n$ are
propagating upstream.
Thus, if the circuit was admitted, the ONU commences the
circuit transmission with the circuit partition of cycle $n+1$.

For signaling packet traffic, the
ONU report in the packet partition of cycle $n-1$ carries
the current queue depth as of the report generation instant.
Based on this queue depth, the OLT determines the effective
bandwidth request and bandwidth allocation as described in
Section~\ref{dba:sec}.
The gate message transmitted downstream at the
beginning of cycle $n$ informs the ONU about its
upstream transmission window in the packet partition of
cycle $n$.

As illustrated in Fig.~\ref{fig:cycle}, in the DyCaPPON design,
the circuit partition is positioned at the beginning of the cycle,
in an effort to mitigate the idle time between the
end of the packet transmissions in the preceding cycle
and the beginning of the packet transmissions of the current cycle.
In particular, when the last packet
transmission of cycle $n-1$ arrives at the OLT at the
end of cycle $n-1$, the first packet transmission of cycle $n$
can arrive at the OLT at the very earliest one roundtrip
propagation delay
(plus typically negligible processing time and gate transmission time)
after the beginning of cycle $n$.
If the circuit partition duration $\Xi(n)$ is longer than the
roundtrip propagation delay $2 \tau$, then idle time between packet
partitions is avoided.
On the other hand, if $\Xi(n) < 2\tau$, then there an idle
channel period of duration $2 \tau - \Xi(n)$ between the end of
the circuit partition and the beginning of the packet
partition in cycle $n$.
\begin{figure*}[t]
\begin{center}
\centering
\setlength{\unitlength}{0.825mm}
\begin{picture}(160,32)
\thicklines
\put(0,20){\line(1,0){160}}
\put(0,30){\line(1,0){160}}
\thinlines
\put(0,5){\line(0,1){25}}
\put(160,5){\line(0,1){25}}
\put(80,10){\line(0,1){20}}

\put(8,20){\line(0,1){10}}
\put(12,20){\line(0,1){10}}
\put(30,20){\line(0,1){10}}
\put(34,20){\line(0,1){10}}
\put(76,20){\line(0,1){10}}
\put(80,20){\line(0,1){10}}

\put(92,20){\line(0,1){10}}
\put(96,20){\line(0,1){10}}
\put(110,20){\line(0,1){10}}
\put(114,20){\line(0,1){10}}
\put(142,20){\line(0,1){10}}
\put(156,20){\line(0,1){10}}

\put(4,25){\makebox(0,0)[]{ $G_1^c$}}
\put(9.5,25){\makebox(0,0)[]{ $t_g$}}
\put(21,25){\makebox(0,0)[]{ $G_5^c$}}
\put(31.5,25){\makebox(0,0)[]{ $t_g$}}
\put(50,25){\makebox(0,0)[]{ $G_{12}^c$}}
\put(77.5,25){\makebox(0,0)[]{ $t_g$}}

\put(85,25){\makebox(0,0)[]{ $G_1^p$}}
\put(93.5,25){\makebox(0,0)[]{ $t_g$}}
\put(102,25){\makebox(0,0)[]{ $G_{2}^p$}}
\put(111.5,25){\makebox(0,0)[]{ $t_g$}}
\put(126,25){\makebox(0,0)[]{ $\cdots$}}
\put(148,25){\makebox(0,0)[]{ $G_{J}^p$}}
\put(157.5,25){\makebox(0,0)[]{ $t_g$}}

\put(20,15){\vector(-1,0){20}}
\put(60,15){\vector(1,0){20}}
\put(40,16){\makebox(0,0)[]{$\Xi(n) + 3t_g$}}
\put(40,11){\makebox(0,0)[]{Circuit partition}}

\put(100,15){\vector(-1,0){20}}
\put(140,15){\vector(1,0){20}}
\put(120,15){\makebox(0,0)[]{Packet partition}}

\put(60,5){\vector(-1,0){60}}
\put(100,5){\vector(1,0){60}}
\put(80,5){\makebox(0,0)[]{Cycle duration $\Gamma$ }}
\end{picture}

\end{center}
\caption{Detailed example illustration of an upstream transmission cycle $n$:
ONUs 1, 5, and 12 have active circuits with bandwidths resulting in
circuit grant durations $G_1^c$, $G_5^c$, and $G_{12}^c$. Each of the $J$
ONUs is allocated a packet grant of duration $G_j^p$ according
to the dynamic packet bandwidth allocation based on the
reported packet traffic; the packet grant accommodates
at least the ONU report (even if there is not payload packet traffic).}
\label{fig:cycle_det}
\end{figure*}
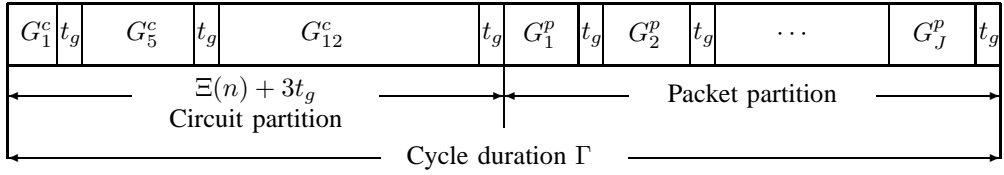

Note that this DyCaPPON design trades off lower responsiveness to
circuit requests for the masking of the roundtrip propagation delay.
Specifically, when an ONU signals a dynamic circuit request in the report
message in cycle $n-1$, it can at the earliest transmit
circuit traffic in cycle $n+1$.
On the other hand, packet traffic signaled in the report message
in cycle $n-1$ can be transmitted in the next cycle, i.e., cycle $n$.

Fig.~\ref{fig:cycle_det} illustrates the
structure of a given cycle in more detail, including
the overheads for the upstream transmissions.
Each ONU that has an active circuit in the cycle
requires one guard time of duration $t_g$ in the
circuit partition.
Thus, with $\eta$ denoting the number of ONUs with active circuits
in the cycle, the duration of the circuit partition is
$\Xi(n) + \eta t_g$.
In the packet partition, each of the $J$ ONUs transmits
at least a report message plus possibly some data upstream,
resulting in an overhead of $J(t_R + t_g)$.
Thus, the overhead per cycle is
\begin{eqnarray}
\omega_o = \eta t_g + J(t_R + t_g).
\end{eqnarray}
The resulting aggregate limit of the transmission windows for packets
in cycle $n$ is
\begin{eqnarray} \label{Gp:eqn}
G^p(n) = \Gamma - \max\{ 2 \tau,\ \Xi(n) \} - \omega_o.
\end{eqnarray}

\subsubsection{Low-Packet-Traffic Mode Polling}
\label{lowload:sec}
\begin{figure*}[t]
\begin{center}
\centering
\setlength{\unitlength}{0.825mm}
\begin{picture}(160,32)
\thicklines
\put(0,0){\line(1,0){140}}
\put(0,20){\line(1,0){140}}
\put(0,21){\makebox(0,0)[b]{OLT}}
\put(0,-1){\makebox(0,0)[t]{ONU}}

\thinlines
\put(8,20){ \vector(1,-3){6.6}}
\thicklines
\put(3,0){\vector(1,3){6.6}}
\put(17,0){\vector(1,3){6.6}}

\thinlines
\put(38,20){\vector(1,-3){6.6}}
\thicklines
\put(31,0){\vector(1,3){6.6}}
\put(45,0){\vector(1,3){6.6}}

\thinlines
\put(67,20){\vector(1,-3){6.6}}
\thicklines
\put(60,0){\vector(1,3){6.6}}
\put(74,0){\vector(1,3){6.6}}

\thinlines
\put(97,20){\vector(1,-3){6.6}}
\thicklines
\put(90,0){\vector(1,3){6.6}}
\put(104,0){\vector(1,3){6.6}}
\put(119,0){\vector(1,3){6.6}}
\put(133,0){\vector(1,3){6.6}}

\put(10,0.75){\makebox(0,0)[b]{$\Xi(n)$}}
\put(143,0.75){\makebox(0,0)[b]{$\Xi(n+1)$}}

\put(26,0.75){\makebox(0,0)[b]{$P1$}}
\put(53,0.75){\makebox(0,0)[b]{$P2$}}
\put(82,0.75){\makebox(0,0)[b]{$P3$}}
\put(112,0.75){\makebox(0,0)[b]{$P4$}}
\put(126,0.75){\makebox(0,0)[b]{Idle}}

\thinlines
\put(10,20){\line(0,1){10}}
\put(140,20){\line(0,1){10}}

\put(3,11){\makebox(0,0)[]{ $R$}}
\put(31,11){\makebox(0,0)[]{ $R$}}
\put(60,11){\makebox(0,0)[]{ $R$}}
\put(90,11){\makebox(0,0)[]{ $R$}}
\put(119,11){\makebox(0,0)[]{ $R$}}

\put(14,11){\makebox(0,0)[]{ $G$}}
\put(43,11){\makebox(0,0)[]{ $G$}}
\put(72,11){\makebox(0,0)[]{ $G$}}
\put(102,11){\makebox(0,0)[]{ $G$}}

\put(50,25){\vector(-1,0){40}}
\put(100,25){\vector(1,0){40}}
\put(76,25){\makebox(0,0)[]{cycle $n$, fixed duration $\Gamma$}}
\end{picture}
\end{center}
\caption{Illustration of low-packet-traffic mode polling:
If transmissions from all ONUs in the packet phase P1
following the circuit partition $\Xi(n)$ reach the OLT more than
$2 \tau$ before the end of the cycle, the OLT can launch
additional packet polling rounds P2, P3, and P4 to serve
newly arrived packet traffic before the next circuit partition
$\Xi(n+1)$.}
\label{fig:cyclell}
\end{figure*}
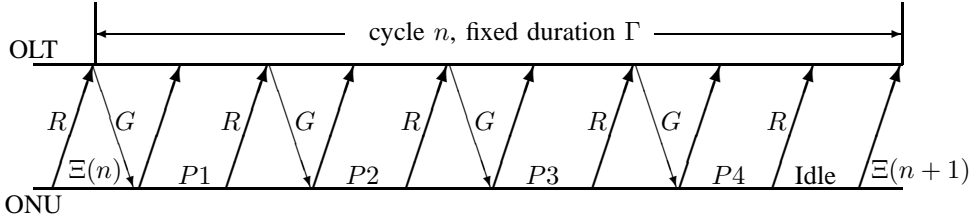

If there is little packet traffic, the circuit partition $\Xi(n)$
and the immediately following packet transmission phase denoted
P1 in Fig.~\ref{fig:cyclell} may leave significant portions of
the fixed-duration cycle idle.
In such low-packet-traffic cycles, the OLT can launch
additional polling rounds denoted P2, P3, and P4 in Fig.~\ref{fig:cyclell}
to serve newly arrived packets with low delay.
Specifically, if all granted packet upstream transmissions have
arrived at the OLT and there is more than
$J(t_R + t_g) + 2 \tau$ time remaining until the end of the cycle
(i.e., the beginning of the arrival of the next circuit
partition $\Xi_{n+1}$) at the OLT, then the OLT can launch
another polling round.

\subsection{Dynamic Circuit Admission Control}
\label{ac:sec}
For each circuit class $k,\ k = 1, 2, \ldots, K$, the OLT tracks the number
$n_k$ of currently active circuits,
i.e., the OLT tracks the state vector $\mathbf{n}:=(n_1,...,n_k)$
representing the numbers of active circuits.
Taking the inner product of $\mathbf{n}$
with the vector $\mathbf{b}:=(b_1,...,b_k)$ representing
the bit rates of the circuit classes gives
the currently required aggregate circuit bandwidth
\begin{eqnarray}
\beta =  \mathbf{b} \cdot \mathbf{n} = \sum_{k=1}^K b_k n_k,
\end{eqnarray}
which corresponds to the circuit partition duration
\begin{eqnarray} \label{Xin:eqn}
\Xi(n) = \frac{\beta \Gamma}{C}.
\end{eqnarray}
For a given limit $C_c,\ C_c \leq C$, of bandwidth available for
circuit service, we let
$\mathcal{S}$ denote the state space of the
stochastic knapsack model~\cite{Ross95} of the dynamic circuits, i.e.,
\begin{eqnarray}
\mathcal{S} := \{\mathbf{n}\in I^{K} : \mathbf {b} \cdot \mathbf {n} \leq C_c \},
\end{eqnarray}
where $I$ is the set of non-negative integers.

For an incoming ONU request for a circuit of class $k$, we let
$\mathcal{S}_k$ denote the subset of the state space $\mathcal{S}$
that can accommodate the circuit request, i.e.,
has at least spare bandwidth $b_k$ before reaching
the circuit bandwidth limit $C_c$. Formally,
\begin{eqnarray}
\mathcal{S}_k:= \{\mathbf{n} \in \mathcal{S} :
            \mathbf {b} \cdot \mathbf {n} \leq C_c - b_k \}.
\end{eqnarray}
Thus, if presently $\mathbf{n} \in \mathcal{S}_k$, then
the new class $k$ circuit can be admitted; otherwise, the
class $k$ circuit request must be rejected (blocked).

\subsection{Packet Traffic Dynamic Bandwidth Allocation}
\label{dba:sec}
With the offline scheduling approach~\cite{ZhMo09} of DyCaPPON,
the reported packet queue occupancy corresponds to the
duration of the upstream packet transmission windows
$R_j, j = 1, 2, \ldots, J$, requested by ONU $j$.
Based on these requests, and the available aggregate
packet upstream transmission window $G^p$ (\ref{Gp:eqn}),
the OLT allocates upstream packet transmission windows with durations
$G^p_j,\ j = 1, 2, \ldots, J$, to the individual ONUs.

The problem of fairly allocating bandwidth so as to enforce a
maximum cycle duration has been extensively studied for the Limited
grant sizing approach~\cite{AsYDA03,BSA06},
which we adapt as follows. We set the
packet grant limit for cycle $n$ to
\begin{eqnarray}
G_{\max}(n) = \frac{G^p(n)}{J}.
\end{eqnarray}
If an ONU requests less than the maximum packet grant
duration $G_{\max}(n)$, it is granted its full request and the
excess bandwidth (i.e., difference between $G_{\max}(n)$ and allocated
grant) is collected by an excess bandwidth distribution
mechanism.
If an ONU requests a grant duration longer than $G_{\max}(n)$, it is allocated
this maximum grant duration, plus a portion of the excess
bandwidth according to the equitable distribution approach
with a controlled excess allocation bound~\cite{AYDA1103,BSA06}.

With the Limited grant sizing approach, there is commonly an unused
slot remainder of the grant allocation to
ONUs~\cite{KrMD2002,HaSM06,NaM06}
due to the next queued packet not fitting into the
remaining granted transmission window.
We model this unused slot remainder by half of the average
packet size $\bar{P}$ for each of the $J$ ONUs.
Thus, the total mean unused transmission window duration
in a given cycle is
\begin{eqnarray}  \label{omegau:eqn}
\omega_u = \frac{J \bar{P}}{2 C}.
\end{eqnarray}

\section{Performance Analysis}
\label{sec:analysis}

\subsection{Circuit Traffic}
\label{percir:sec}
\subsubsection{Request Blocking}
\label{reqblo:sec}
In this section, we employ techniques from the analysis of
stochastic knapsacks~\cite{Ross95} to evaluate the blocking probabilities
$B_k$ of the circuit class.
We also evaluate the mean duration of the circuit partition $\Xi$,
which governs the mean available packet partition duration $G^p$,
which in turn is a key parameter for the evaluation of the mean
packet delay in Section~\ref{delan:sec}.

The stochastic knapsack model~\cite{Ross95} is a generalization of the
well-known Erlang loss system model to circuits with heterogeneous
bandwidths.
In brief, in the stochastic knapsack model, objects of different
classes (sizes) arrive to a knapsack of fixed capacity (size)
according to a stochastic arrival process.
If a newly arriving object fits into the currently vacant knapsack
space, it is admitted to the knapsack and remains in the knapsack
for some random holding time. After the expiration of the holding time,
the object leaves the knapsack and frees up the knapsack space that
it occupied.
If the size of a newly arriving object exceeds the currently
vacant knapsack space, the object is blocked from entering the knapsack,
and is considered dropped (lost).

We model the prescribed limit $C_c$ on the bandwidth available for circuit
service as the knapsack capacity.
The requests for circuits of bandwidth $b_k,\ k = 1, 2, \ldots, K$,
arriving according to a Poisson process with rate
$p_k \lambda_c$ are modeled as the objects seeking entry into the knapsack.
An admitted circuit of class $k$ occupies the bandwidth (knapsack space)
$b_k$ for an exponentially distributed holding time with mean $1/\mu$.

We denote $\mathcal{S}(\beta)$ for the set of states
$\mathbf{n}$ that occupy an aggregate bandwidth
$\beta,\ 0 \leq \beta \leq C_c$, i.e.,
\begin{eqnarray}
\mathcal{S}(\beta) := \{\mathbf{n}\in \mathcal{S} :
        \mathbf {b} \cdot \mathbf {n}  = \beta \}.
\end{eqnarray}
Let $q(\beta)$ denote the equilibrium probability of the
currently active circuits occupying an aggregate bandwidth of $\beta$.
Through the recursive Kaufman-Roberts algorithm~\cite[p. 23]{Ross95},
which is given in the Appendix, the equilibrium probabilities
$q(\beta)$ can be computed with a time complexity of
$O(C_c K)$ and a memory complexity of $O(C_c+K)$.

The blocking probability $B_k,\ k = 1, 2, \ldots, K$ is obtained by summing
the equilibrium probabilities $q(\beta)$ of the sets of states
that have less than $b_k$ available circuit bandwidth, i.e.,
\begin{eqnarray}  \label{Bk:eqn}
B_k =  \sum_{\beta = C_c - b_k + 1}^{C_c} q(\beta).
\end{eqnarray}
We define the average circuit blocking probability
\begin{eqnarray}
\bar{B} =  \sum_{k = 1}^K p_k B_k.
\end{eqnarray}

\subsubsection{Aggregate Circuit Bandwidth}
The performance evaluation for packet delay in
Section~\ref{pkt_perf:sec} requires taking
expectations over the distribution $q(\beta)$
of the aggregate bandwidth $\beta$ occupied by circuits.
In preparation for these packet evaluations, we define
$\E_{\beta}[f(\beta)]$ to denote the expectation of
a function $f$ of the random variable $\beta$ over the
distribution $q(\beta)$, i.e., we define
\begin{eqnarray}  \label{Ebeta:eqn}
\E_{\beta}[f(\beta)] =  \sum_{\beta = 0}^{C_c} f(\beta) q(\beta).
\end{eqnarray}
With this definition, the mean aggregate bandwidth of the active circuits is
obtained as
\begin{eqnarray}  \label{beta_avg:eqn}
  \bar{\beta} =
\E_{\beta}[\beta] =  \sum_{\beta = 0}^{C_c} \beta q(\beta).
\end{eqnarray}
Note that by taking the expectation of (\ref{Xin:eqn}), the corresponding
mean duration of the circuit partition is
$   \bar{\Xi} =
\E_{\beta}[\beta \Gamma /C] = \bar{\beta} \Gamma /C$.

\subsubsection{Delay and Delay Variation}
\label{cirdel:sec}
In this section we analyze the delay and delay variations
experienced by circuit traffic as it traverses a DyCaPPON network
from ONU to OLT.
Initially we ignore delay variations, i.e., we consider
that a given circuit with bit rate $b$ has a fixed position for
the transmission of its $b \Gamma$ bits in each cycle.
Three delay components arise:
The ``accumulation/dispersal'' delay of $\Gamma$ for the
$b \Gamma$ bits of circuit traffic that are transmitted per cycle.
Note that the first bit arriving to form a ``chunk'' of
$b \Gamma$ bits experiences the delay $\Gamma$ at the ONU, waiting
for subsequent bits to ``fill up (accumulate)'' the chunk.
The last bit of a chunk experiences essentially no delay at the ONU, but
has to wait for a duration of $\Gamma$ at the OLT to ``send out (disperse)''
the chunk at the circuit bit rate $b$.
The other delay components are the transmission delay of $b \Gamma/C$
and the propagation delay $\tau$.
Thus, the total delay is
\begin{eqnarray}
\Gamma \left( 1 + \frac{b}{C} \right) + \tau.
\end{eqnarray}

Circuit traffic does not experience delay variations (jitter) in
DyCaPPON as long as the positions (in time) of the circuit transmissions
in the cycle are held fixed.
When an ongoing circuit is closing down or a new
circuit is established, it may become necessary to
rearrange the transmission positions of the circuits in the cycle in
order to keep all circuit transmissions within the circuit partition
at the beginning of the cycle and avoid idle times during
the circuit partition.
Adaptations of packing algorithms~\cite{dyck90}
could be employed to minimize the shifts in transmission positions.
Note that for a given circuit service limit $C_c$,
the worst-case delay variation for a given circuit with rate $b$ is less than
$\Gamma(C_c - b)/C$ as the circuit could at the most shift from
the beginning to the end of the circuit partition of
maximum duration $\Gamma C_c/C$.

\subsection{Packet Traffic}
\label{pkt_perf:sec}
\subsubsection{Stability Limit}
\label{pastab:sec}
Inserting the circuit partition duration $\Xi$ from (\ref{Xin:eqn})
into the expression for the aggregate limit $G^p$ on the
transmission window for packets in a cycle from (\ref{Gp:eqn})
and taking the expectation $\E_{\beta}[\cdot]$ with respect to the distribution
of the aggregate circuit bandwidth $\beta$, we obtain
\begin{eqnarray}   \label{Gp_avg:eqn}
\bar{G^p} = \Gamma -
   \E_{\beta} \left[\max \left\{ 2 \tau,\
             \frac{\beta \Gamma}{C} \right\} \right] - \omega_o.
\end{eqnarray}
Considering the unused slot remainder $\omega_u$ (\ref{omegau:eqn}),
the mean portion of a cycle available for
upstream packet traffic transmissions is limited to
\begin{eqnarray}  \label{pimax:eqn}
\pi_{\max} = 1
    - E_{\beta}\left[ \max \left\{ \frac{2 \tau}{\Gamma},\
    \frac{\beta}{C} \right\} \right]
    - \frac{\omega_o + \omega_u}{\Gamma}.
\end{eqnarray}
That is, the packet traffic intensity $\pi$ must be less than
$\pi_{\max}$ for stability of the packet service, i.e.,
for finite packet delays.

\subsubsection{Mean Delay}
\label{delan:sec}
In this section, we present for stable packet service
an approximate analysis of the mean delay $D$ of packets
transmitted during the packet partition.
In DyCaPPON, packets are transmitted on the bandwidth
that is presently not occupied by admitted circuits.
Thus, fluctuations in the aggregate occupied circuit bandwidth $\beta$
affect the packet delays.
If the circuit bandwidth $\beta$ is presently high,
packets experience longer delays than for presently low
circuit bandwidth $\beta$.
The aggregated occupied circuit bandwidth $\beta$ fluctuates
as circuits are newly admitted and occupy bandwidth and as
existing circuits reach the end of their holding time and release
their occupied bandwidth.
The time scale of these fluctuations of $\beta$ increases
as the average circuit holding time $1/\mu$ increases,
i.e., as the circuit departure rate $\mu$ decreases
(and correspondingly, the circuit request arrival rate $\lambda$ decreases
for a given fixed circuit traffic load $\chi$)~\cite{gaver82}.

For circuit holding times that are orders of magnitude larger than
the typically packet delays (service times) in the system, the
fluctuations of the circuit bandwidth $\beta$ occur at a
significantly longer (slower) time scale than the packet service
time scale. That is, the bandwidth $\beta$ occupied by circuits
exhibits significant correlations over time which in turn give rise
to complex correlations with the packet queueing
delay~\cite{wein80,tham83}. For instance, packets arriving during a
long period of high circuit bandwidth may experience very long
queueing delays and are possibly only served after some circuits
release their bandwidth. As illustrated in
Section~\ref{mu_impact:sec}, the effects of these complex
correlations become significant for scenarios with moderate to long
circuit holding times $1/\mu$ when the circuit traffic load is low
to moderate relative to the circuit bandwidth limit $C_c$ (so that
pronounced circuit bandwidth fluctuations are possible), and the
packet traffic load on the remaining bandwidth of approximately $C -
C_c$ is relatively high, so that substantial packet queue build-up
can occur. We leave a detailed mathematical analysis of the complex
correlations occurring in these scenarios in the context of DyCaPPON
for future research.

In the present study, we focus on an approximate
packet delay analysis that neglects the outlined correlations.
We base our approximate packet delay analysis
on the expectation $E_{\beta}[f(\beta)]$ (\ref{Ebeta:eqn}),
i.e., we linearly weigh packet delay metrics $f(\beta)$
with the probability masses $q(\beta)$ for the aggregate
circuit bandwidth $\beta$.
We also neglect the ``low-load'' operating mode of
Section~\ref{lowload:sec} in the analysis.

In the proposed DyCaPPON cycle structure, a packet experiences five
main components, namely $(i)$ the reporting delay from the
generation instant of the packet to the transmission of the report
message informing the OLT about the packet,
which for the fixed cycle duration of DyCaPPON
equals half the cycle duration, i.e., $\Gamma/2$,
$(ii)$ the report-to-packet partition delay $D_{\mathrm{r-p}}$
from the instant of report
transmission to the beginning of the packet partition in the next
cycle,
$(iii)$ the queuing delay $D_q$ from the reception instant of the
grant message to the beginning of the transmission of the packet, as
well as $(iv)$ the packet transmission delay with mean $\bar{P}/C$,
and $(v)$ the upstream propagation delay $\tau$.

In the report-to-packet partition delay we include a delay component
of half the mean duration of the packet partition $\bar{G^p}/2$
to account for the
delay of the reporting of a particular ONU to the end of the packet partition.
The delay from the end of the packet partition in one cycle
to the beginning of the packet partition of the next cycle is the
maximum of the roundtrip propagation delay $2 \tau$ and the mean
duration of the circuit partition $\Xi$.
Thus, we obtain overall for the report-to-packet partition delay
\begin{eqnarray}
D_{\mathrm{r-p}}
     &=& \frac{\bar{G^p}}{2}
    + \E_{\beta} \left[ \max \left\{2 \tau,\
                \frac{\beta \Gamma}{C}  \right\} \right]\\
              &=& \frac{1}{2}
        \left( \Gamma +  \E_{\beta} \left[\max \left\{  2 \tau,\
                     \frac{\beta \Gamma}{C} \right\} \right]
        - \omega_o \right).
\end{eqnarray}

We model the queueing delay with an M/G/1 queue.
Generally, for messages with mean service time $\bar{L}/C$,
normalized message size variance $\sigma^2/\bar{L}^2$, and
traffic intensity $\rho$,
the M/G/1 queue has expected queueing delay~\cite{Kleinrock75}
\begin{eqnarray}
D_{M/G/1} = \frac{\rho \frac{\bar{L}}{C}
 \left( 1+\frac{\sigma^2}{\bar{L}^2} \right) }{2(1-\rho)}.
\end{eqnarray}
For DyCaPPON, we model the aggregate packet traffic from all $J$
ONUs as feeding into one M/G/1 queue with mean packet size $\bar{P}$
and packet size variance $\sigma^2_p$.
We model the circuit partitions, when the
upstream channel is not serving packet traffic, through scaling of the
packet traffic intensity.
In particular, the upstream channel is available for
serving packet traffic only for the mean fraction
$(\bar{G^p} - \omega_u) / \Gamma$ of a cycle.
Thus, for large backlogs served across several cycles,
the packet traffic intensity during the packet partition is effectively
\begin{eqnarray}
\pi_{\rm eff}  = \frac{\pi}{\pi_{\max}}.
\end{eqnarray}
Hence, the mean queueing delay is approximately
\begin{eqnarray}
D_{q} &=&
          \frac{ \frac{ \pi_{\rm eff} \bar{P} }{C}
           \left( 1 + \frac{\sigma_p^2}{\bar{P}^2} \right)}
  {2 ( 1- \pi_{\rm eff} ) }.
\end{eqnarray}
Thus, the overall mean packet delay is approximately
\begin{eqnarray}
D = \frac{\Gamma}{2} + D_{\mathrm{r-p}} +  D_q + \frac{\Bar{P}}{C} + \tau.
\end{eqnarray}

\section{DyCaPPON Performance Results}
\label{eval:sec}
\subsection{Evaluation Setup}
\label{eval_setup:sec}
We consider an EPON with $J = 32$ ONUs, a channel bit rate $C = 10$~Gb/s,
and a cycle duration $\Gamma = 2$~ms.
Each ONU has abundant buffer space and a one-way propagation delay
of $\tau = 96~\mu$s to the OLT.
The guard time is $t_g = 5\ \mu$s and the report message has 64 Bytes.
We consider $K = 3$ classes of circuits as specified in Table~\ref{cir:tab}.
\begin{table}[t]
\caption{Circuit bandwidths $b_k$ and request probabilities $p_k$ for $K=3$
classes of circuits in performance evaluations.}
\label{cir:tab}
\vspace{-0.25cm}
\begin{center}
\begin{tabular}{|l|rrr|} \hline
   & \multicolumn{3}{|c|}{Class $k$}\\
   &           1  &   2   &  3   \\ \hline
$b_k$ [Mb/s] & 52 &  156  & 624   \\
$p_k$ [\%]   & 53.56   & 28.88    & 15.56  \\
\hline
\end{tabular}
\end{center}
\end{table}
A packet has 64~Bytes with 60\% probability,
300~Bytes with 4\% probability,  580~Bytes with 11\% probability,
and 1518~bytes with 25\% probability, thus
the mean packet size is $\bar{P}=493.7$~Bytes.
The verifying simulations were conducted with a CSIM based simulator and
are reported with 90~\% confidence intervals which are too
small to be visible in the plots.

\begin{table*}[t]
\caption{Circuit blocking probabilities $B_k$
from analysis (A) Eqn.~(\ref{Bk:eqn}) with representative
verifying simulations (S) for given offered circuit traffic load $\chi$,
circuit bandwidth limit $C_c = 2$ or 4~Gb/s
and mean circuit holding time $1/\mu$.
The blocking probabilities are independent of the packet
traffic load $\pi$.
Table also gives average circuit traffic bit rate $\bar{\beta}$
from (\ref{beta_avg:eqn}),
mean duration of packet phase $\bar{G}_p$ (\ref{Gp_avg:eqn}),
and packet traffic load limit $\pi_{\max}$ (\ref{pimax:eqn}).
}
\label{pi_bk:tab}
\begin{center}
\begin{tabular}{|c|c|l|ccc|c|ccc|} \hline         &&&&&&&&& \\
       ~~~~$\chi$ & $C_c$ & $1/\mu$&  $B_1$  &  $B_2$   &  $B_3$ & $\bar{B}$  &
            $\bar{\beta}$  & $\bar{G}_p$  & $\pi_{\max}$  \\
      & [Gb/s]  & [s] &  [\%] & [\%]     & [\%]      & [\%]  &
            $[10^{9}Gbps]$  & [ms]  &   \\ \hline
        $0.1 $ & 4 & A&  $8.5 \cdot 10^{-3}$ & $ 0.031$ &  0.28
          & $0.057$  & 1.05 & 1.68   & 0.842  \\
      $ 0.1 $ & 2 &  A &  0.93 & 3.2 &  21
          & 4.6  & 0.93 & 1.70  & 0.852  \\
        $0.1 $ & 2 & 0.5 S&  0.72  & 2.9 &  21
          & 4.4  & 0.90 &  &   \\
      $0.1 $ & 2 & 0.02 S&  1.1 & 3.7 &  22
          & 5.1  & 0.95 &  &   \\ \hline
        $0.4 $ & 4 & A&  3.34  &  10.6  &  39.6 & 10.9 &
            3.02  & 1.33  & 0.665  \\
$0.4 $ & 4 & 0.5 S& 3.4 &  11  &  41 & 11 &
            3.0 &  &   \\
         $0.4 $ & 4 & 0.02 S&  4.4 &  12  &  42 & 13 &
            3.2 &   &   \\
      $0.4 $  & 2 & A&  12.1  &  33.1  &  85.7 & 29.6 &
            1.68  & 1.60  & 0.799  \\
         $0.4 $ & 2  &  0.5 S&  12  &  35  &  85 & 30 &
            1.6  &   &   \\
        $0.4 $   & 2 &  0.02  S&  13  &  35  &  87 & 31 &
            1.7  &  &    \\ \hline
         $0.7 $ & 4 & A&  9.55  & 26.5 &  74.6 & 24.6  &
           3.49  & 1.24   & 0.618  \\
 $0.7 $ & 4 & 0.5 S&  10  &  27  &  75 & 25  &
           3.5  &  &   \\
         $0.7 $ & 4 & 0.02 S& 13   &  29  &  75 & 28  &
           3.6  &  &   \\
         $ 0.7 $ & 2 & A&  23.5  &  56.6  &  98.3 & 44.7  &
           1.83  & 1.57  & 0.785 \\
         $0.7 $ & 2 & 0.5 S&  23  &  57 &  98 & 45  &
           1.8 &   &   \\
        $0.7 $   & 2& 0.02 S& 28 & 57  & 98 & 47 &
           1.8  &   &    \\ \hline
\end{tabular}
\end{center}
\end{table*}
\begin{figure}[t]
\begin{center}
\includegraphics[scale=0.6]{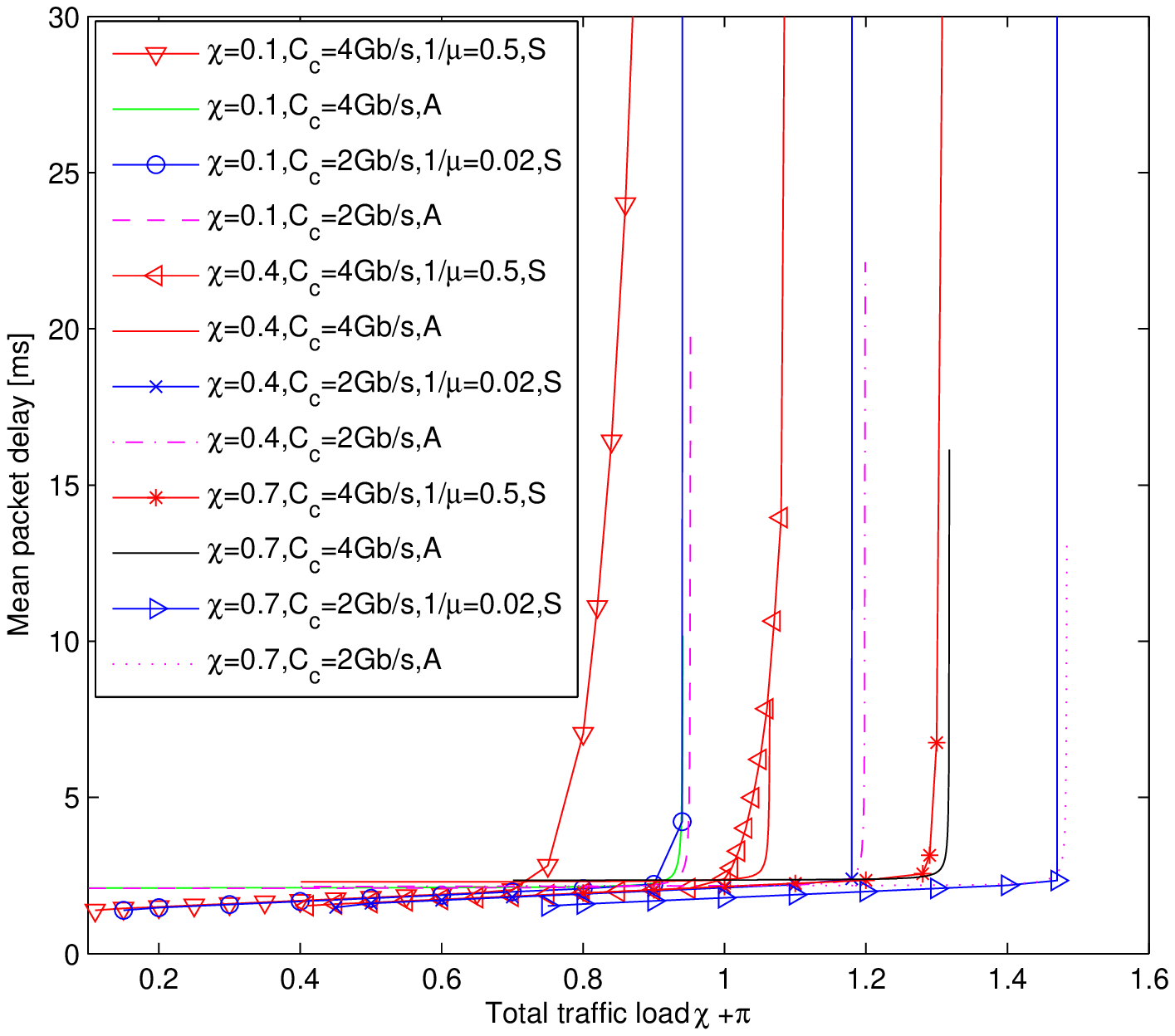}
\end{center}
\caption{Impact of packet traffic load $\pi$:
Mean packet delay $D$ from simulations (S) and
analysis (A) as a function of total traffic load $\chi + \pi$,
which is varied by varying $\pi$ for fixed
circuit traffic load $\chi = 0.1$, 0.4, or 0.7.}
\label{fig:pi}
\end{figure}
\begin{figure}[t]
\begin{center}
\includegraphics[scale=0.675]{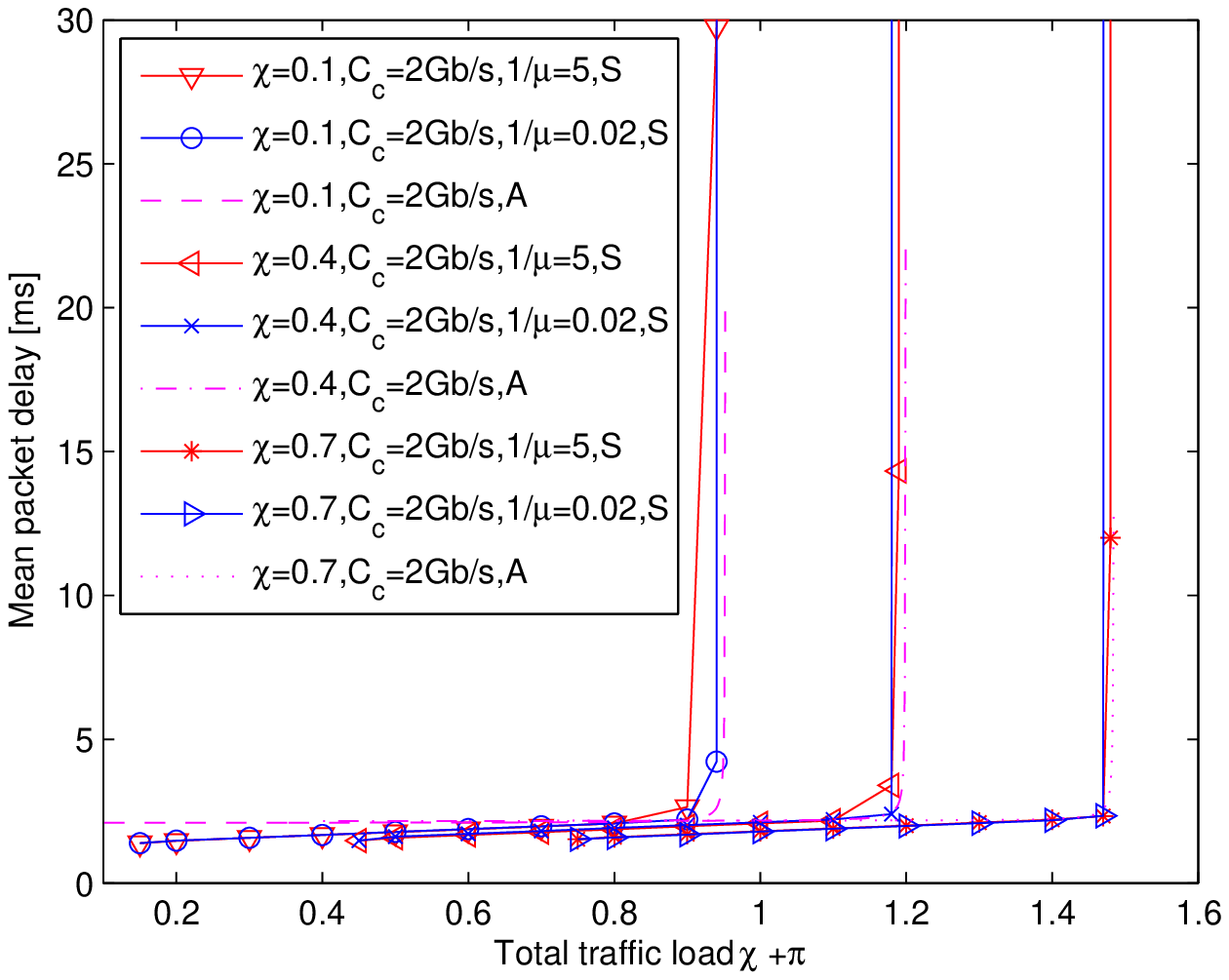}
\end{center}
\caption{Impact of packet traffic load $\pi$:
Mean packet delay $D$ from simulations (S) and
analysis (A) as a function of total traffic load $\chi + \pi$,
which is varied by varying $\pi$ for fixed
circuit traffic load $\chi = 0.1$, 0.4, or 0.7,
with $C_c=2$~Gb/s, and two different $1/\mu$ values.}
\label{fig:pi1}
\end{figure}
\begin{figure}[t]
\begin{center}
\includegraphics[scale=0.675]{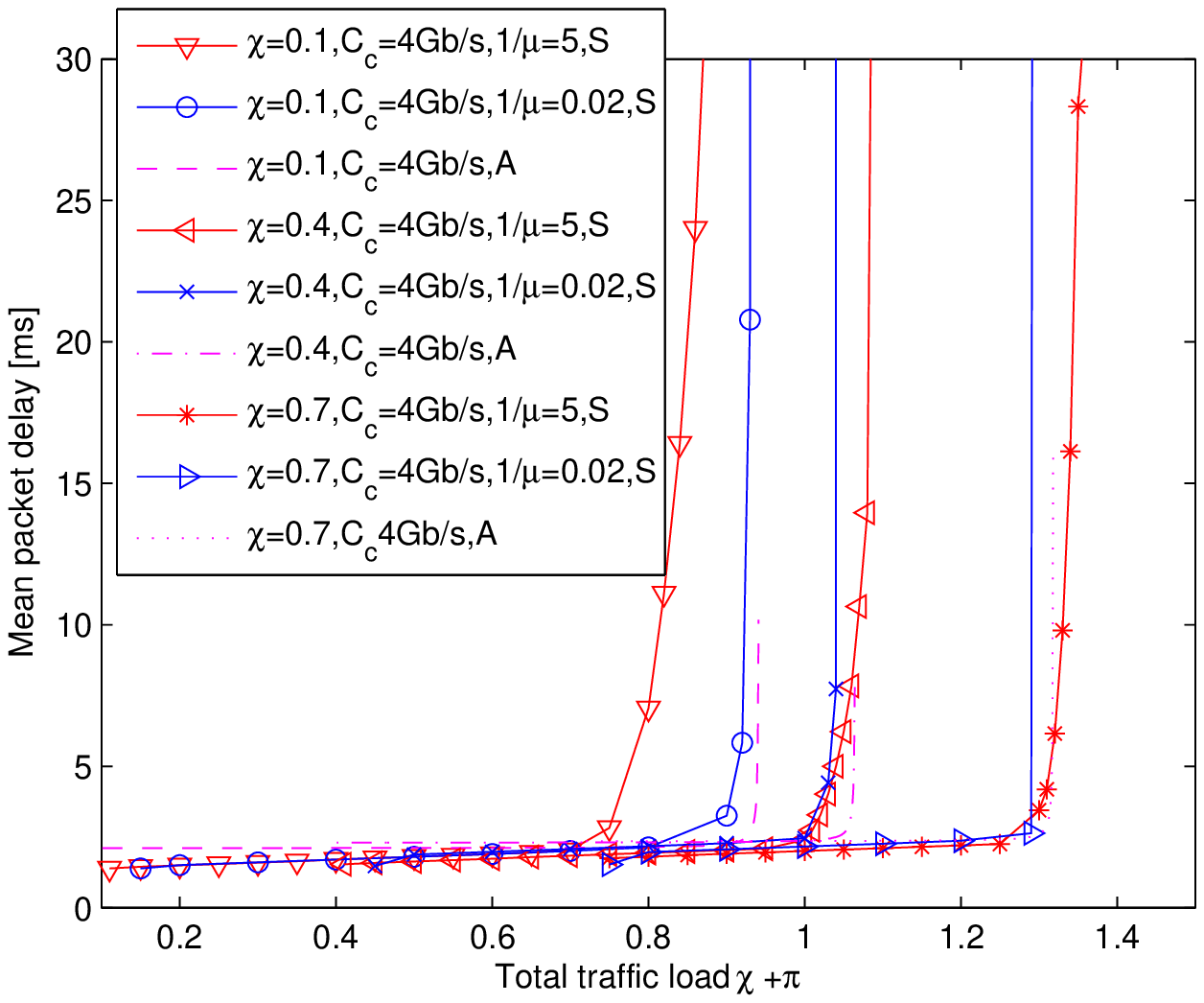}
\end{center}
\caption{Impact of packet traffic load $\pi$:
Mean packet delay $D$ from simulations (S) and
analysis (A) as a function of total traffic load $\chi + \pi$,
which is varied by varying $\pi$ for fixed
circuit traffic load $\chi = 0.1$, 0.4, or 0.7,
with $C_c=4$~Gb/s, and two different $1/\mu$ values.}
\label{fig:pi2}
\end{figure}
\begin{figure}[t]
\begin{center}
\includegraphics[scale=0.675]{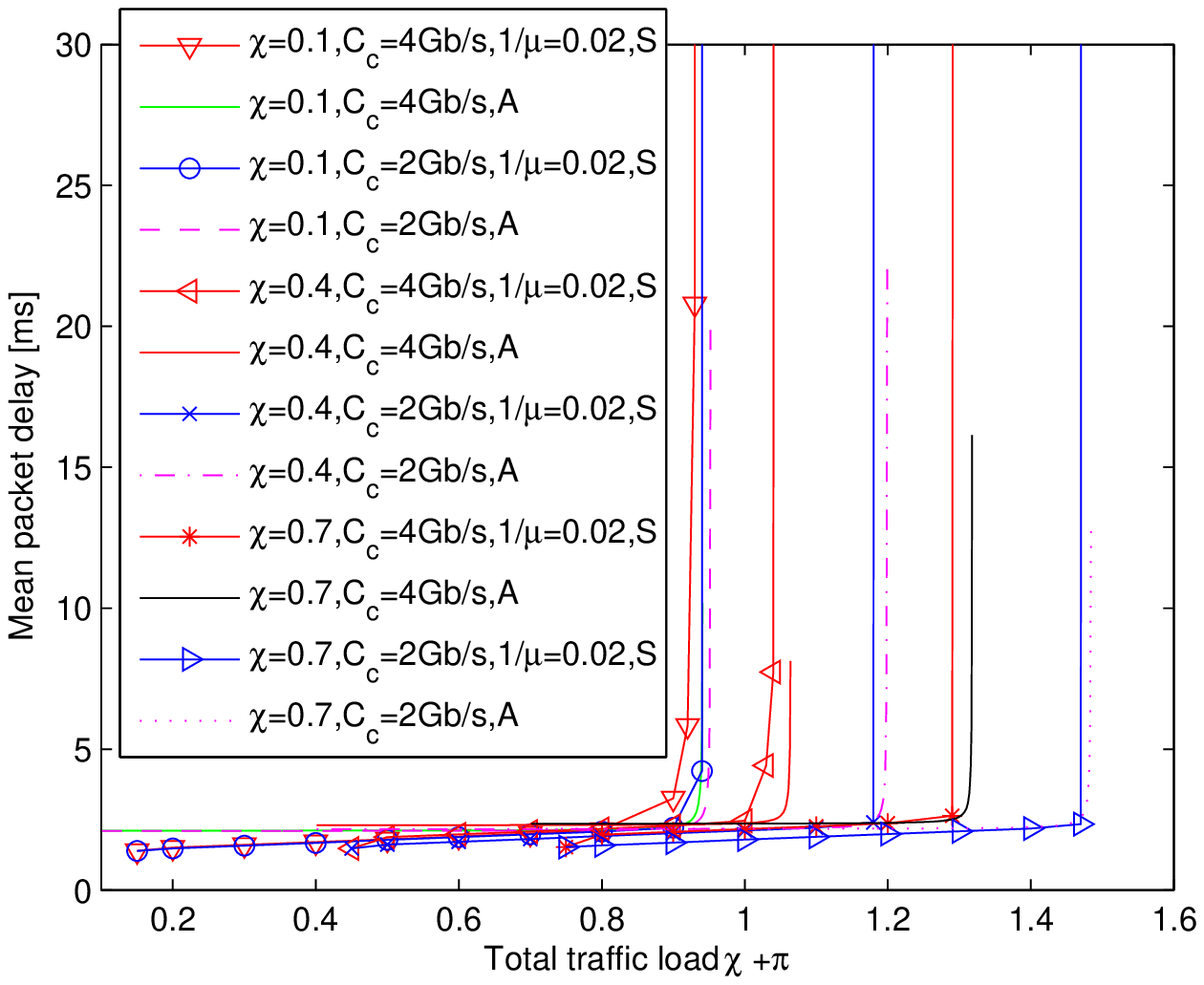}
\end{center}
\caption{Impact of packet traffic load $\pi$:
Mean packet delay $D$ from simulations (S) and
analysis (A) as a function of total traffic load $\chi + \pi$,
which is varied by varying $\pi$ for fixed
circuit traffic load $\chi = 0.1$, 0.4, or 0.7,
with $1/\mu=0.02$~s, and two different $C_c$ values.}
\label{fig:pi3}
\end{figure}
\begin{figure}[t]
\begin{center}
\includegraphics[scale=0.625]{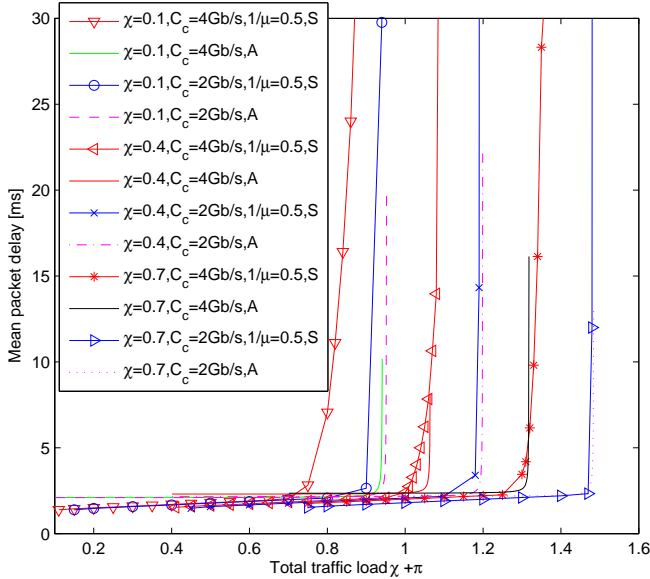}
\end{center}
\caption{Impact of packet traffic load $\pi$:
Mean packet delay $D$ from simulations (S) and
analysis (A) as a function of total traffic load $\chi + \pi$,
which is varied by varying $\pi$ for fixed
circuit traffic load $\chi = 0.1$, 0.4, or 0.7,
with $1/\mu=0.5$~s, and two different $C_c$ values.}
\label{fig:pi4}
\end{figure}
\subsection{Impact of Packet Traffic Load $\pi$}
\label{pi_impact:sec}
In Table~\ref{pi_bk:tab} we present circuit blocking probability results.
In Figs.~\ref{fig:pi}--\ref{fig:pi4} we plot
packet delay results for increasing packet traffic load $\pi$.
We consider three levels of offered circuit traffic load $\chi$,
which are held constant as the packet traffic load $\pi$ increases.
DyCaPPON ensures consistent circuit service with the
blocking probabilities and
delay characterized in Section~\ref{percir:sec} irrespective
of the packet traffic load $\pi$, that is, the packet traffic does
\textit{not} degrade the circuit service at all.
Specifically, Table~\ref{pi_bk:tab} gives
the blocking probabilities $B_k$ as well as the average
circuit blocking probability $\bar{B} = \sum_{k = 1}^K p_k B_k$
for the different levels of offered circuit traffic load;
these blocking probability values hold for the full range
of packet traffic loads $\pi$.

We observe from Table~\ref{pi_bk:tab} that for a given
offered circuit traffic load level $\chi$, the blocking probability increases
with increasing circuit bit rate $b_k$ as it is less likely that sufficient
bit rate is available for a higher bit rate circuit.
Moreover, we observe that the blocking probabilities increase with
increasing offered circuit traffic load $\chi$.
This is because the circuit transmission limit $C_c$ becomes
increasingly saturated with increasing offered circuit load
$\chi$, resulting in more blocked requests.
The representative simulation results in Table~\ref{pi_bk:tab}
indicate that the stochastic knapsack analysis is accurate,
as has been extensively verified
in the context of general circuit switched systems~\cite{Ross95}.

In Fig.~\ref{fig:pi} we plot the mean packet delay as
a function of the total traffic load, i.e., the sum of offered
circuit traffic load $\chi$ plus the packet traffic load $\pi$.
We initially exclude the scenario with $\chi = 0.1$, $C_c = 4$~Gbps,
and $1/\mu = 0.5$~s from consideration; this scenario is discussed in
Section~\ref{mu_impact:sec}.
We observe from Fig.~\ref{fig:pi} that for low packet traffic load $\pi$
(i.e., for a total traffic load $\chi+ \pi$ just above the offered
circuit traffic load $\chi$),
the packet delay is nearly independent of the offered circuit traffic
load $\chi$.
For low packet traffic load, the few packet transmissions
fit easily into the packet partition of the cycle.

We observe from Figs.~\ref{fig:pi}--\ref{fig:pi4}
sharp packet delay increases for
high packet traffic loads $\pi$ that approach the maximum total
traffic load, i.e., offered circuit traffic load $\chi$ plus maximum
packet traffic load $\pi_{\max}$.
For $C_c = 2$~Gb/s, the maximum packet traffic load
$\pi_{\max}$ is 0.85 for $\chi = 0.1$ and 0.78 for $\chi = 0.7$, see
Table~\ref{pi_bk:tab}. Note that the maximum packet traffic load
$\pi_{\max}$ depends on the offered circuit traffic load $\chi$ and
the circuit traffic limit $C_c$. For a low offered circuit traffic
load $\chi$ relative to $C_c/C$, few circuit requests are blocked
and the admitted circuit traffic load (equivalently mean aggregate
circuit bandwidth $\bar{\beta}$) is close to the offered circuit
load $\chi$. On the other hand, for high offered circuit traffic
load $\chi$, many circuit requests are blocked, resulting in an
admitted circuit traffic load (mean aggregate circuit bandwidth
$\bar{\beta}$) significantly below the offered circuit traffic load
$\chi$. Thus, the total (normalized) traffic load, i.e., offered
circuit load $\chi$ plus packet traffic load $\pi$,
in a stable network can exceed one for
high offered circuit traffic load $\chi$.

\subsection{Impact of Mean Circuit Holding Time}
\label{mu_impact:sec}
\begin{figure}[t]
\includegraphics[scale=0.675]{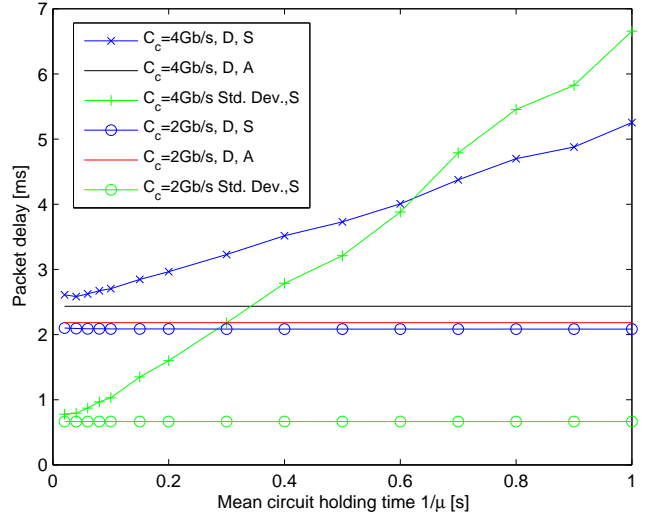}
\caption{Mean packet delay $D$ and standard deviation of
packet delay as a function of mean circuit
holding time $1/\mu$; fixed parameters
$\chi = 0.5$, $\pi = 0.6$.}
\label{fig:mu}
\end{figure}
We now turn to the packet delay results for the
scenario with low circuit traffic load $\chi = 0.1$ relative to the
circuit bandwidth limit $C_c = 4$~Gbps and moderately long
mean circuit holding time $1/\mu = 0.5$~s,
which is included in Figs.~\ref{fig:pi} and~\ref{fig:pi4}.
We observe for this scenario that the mean packet delays
obtained from the simulations begin to increase dramatically
as the total load $\chi+\pi$ approaches 0.8.
In contrast, for the circuit traffic load $\chi = 0.1$
in conjunction with the lower circuit bandwidth limit $C_c = 2$~Gbps
and short mean circuit holding times $1/\mu = 0.02$~s,
the mean packet delays remain low for total loads up to
close to the total maximum load $\chi + \pi_{\max} = 0.95$
and then increase sharply.

The pronounced delay increases at lower loads (in the 0.75--0.92
range) for the $\chi = 0.1$, $C_c = 4$~Gbps, $1/\mu = 0.5$~s
scenario are mainly due to the higher-order complex correlations
between the pronounced slow-time scale fluctuations of the
circuit bandwidth and the packet queueing
as explained in Section~\ref{delan:sec}.
The high circuit bandwidth limit $C_c = 4$~Gbps relative to the
low circuit traffic load $\chi = 0.1$ allows
pronounced fluctuations of the aggregate occupied circuit bandwidth
$\beta$. For the moderately long mean circuit holding time $1/\mu =
0.5$~s, these pronounced fluctuations occur at a long time scale
relative to the packet service time scales, giving rise to
pronounced correlation effects. That is, packets arriving during
periods of high circuit bandwidth $\beta$ may need to wait (queue)
until some circuits end and release sufficient bandwidth to serve
the queued packet backlog. These correlation effects are neglected
in our approximate packet delay analysis in Section~\ref{delan:sec}
giving rise to the large discrepancy between simulation and analysis
observed for the $\chi = 0.1$, $C_c = 4$~Gb/s, $1/\mu = 0.5$~s scenario in
Fig.~\ref{fig:pi}.

We observe from Fig.~\ref{fig:pi} for the scenarios with
relatively high circuit traffic loads $\chi = 0.4$ and~0.7
relative to the considered circuit bandwidth limits $C_c = 2$
and 4~Gbps that the mean packet delays remain low up to
levels of the total load close to the total stability limit
$\chi + \pi_{\max}$ predicted from the stability analysis
in Section~\ref{pastab:sec}.
The relatively high circuit traffic loads $\chi$
lead to high circuit blocking probabilities
(see Table~\ref{pi_bk:tab}) and the admitted circuits utilize the
available circuit traffic bandwidth $C_c$ nearly fully
for most of the time. Vacant portions of the
circuit bandwidth $C_c$ are quickly occupied by the frequently arriving
new circuit requests. Thus, there are only relatively minor
fluctuations of the bandwidth available for packet service
and the approximate packet delay analysis is quite accurate.

Returning to the scenario with relatively low circuit traffic load
$\chi = 0.1$ in Fig.~\ref{fig:pi}, we observe that for
the short mean circuit holding time $1/\mu = 0.02$, the
mean packet delays remain low up to load levels close to the
stability limit $\chi + \pi_{\max}$.
For these relatively short circuit durations, the pronounced
fluctuations of the occupied circuit bandwidth occur on a sufficiently
short time scale to avoid significant higher-order correlations between
the circuit bandwidth and the packet service.

We examine these effects in more detail in
Fig.~\ref{fig:mu}, which shows means and standard deviations of
packet delays as a function of the mean circuit holding time $1/\mu$
for fixed traffic load $\chi = 0.5$, $\pi = 0.6$. We observe that
for the high $C_c = 4$~Gbps circuit bandwidth limit, the mean packet
delay as well as the standard deviation of the packet delay obtained
from simulations increase approximately linearly with increasing
mean circuit holding time $1/\mu$. The $C_c = 4$~Gbps circuit
bandwidth limit permits sufficiently large fluctuations of the
circuit bandwidth $\beta$ for the $\chi = 0.5$ load, such that for
increasing circuit holding time, the packets increasingly experience
large backlogs that can only be cleared when some circuits end and
release their bandwidth. In contrast,
for the lower circuit bandwidth limit $C_C = 2$~Gbps, which severely
limits fluctuations of the circuit bandwidth $\beta$ for the high
circuit traffic load $\chi = 0.5$, the mean and standard deviation
of the packet delay remain essentially constant for increasing
$1/\mu$.

\begin{table}[t]
\caption{Mean circuit blocking probability $\bar{B}$
and mean packet delay $D$
as a function of circuit traffic load $\chi$;
fixed parameters: circuit bandwidth limit $C_c = 2$~Gb/s,
packet traffic load $\pi = 0.7$.}
\label{fig:chi}
\vspace{-0.75cm}
\begin{center}
\begin{tabular}{|l|ccccccc|} \hline
$\chi$ &  $0.0001$ &  $0.05$  &   $0.1$   &  $0.20$
        & $0.40$ & $0.60$ & $ \chi \rightarrow \infty$  \\ \hline
$\bar{B}$, S [\%] & 0 & 1.2  &  5.1  &  16 & 31 & 43  & \\
$\bar{B}$, A [\%] & 0.016 & 1.08  &  4.81 &  14.9 & 29.6  & 40.1 &100\\ \hline
$D$, S [ms] & 1.9 &  2.0  & 2.0  &  2.1 & 2.2 &  2.2&  \\
$D$, A [ms] & 2.10&  2.11 & 2.13 & 2.16 & 2.21& 2.23& 2.42  \\   \hline
\end{tabular}
\end{center}
\end{table}
\subsection{Impact of Offered Circuit Traffic Load $\chi$}
\label{chi_impact:sec}
In Table.~\ref{fig:chi}, we examine the impact of the
circuit traffic load $\chi$ on the DyCaPPON performance
more closely.
We keep the packet traffic load fixed at $\pi = 0.7$
and examine the average circuit blocking probability $\bar{B}$
and the mean packet delay $D$ as a function of the circuit traffic load $\chi$.
We observe from Table.~\ref{fig:chi} that, as expected, the
mean circuit blocking probability $\bar{B}$ increases with increasing
circuit traffic load $\chi$, whereby analysis closely matches the simulations.

\begin{figure}[t]
\begin{tabular}{c}
\includegraphics[scale=0.615]{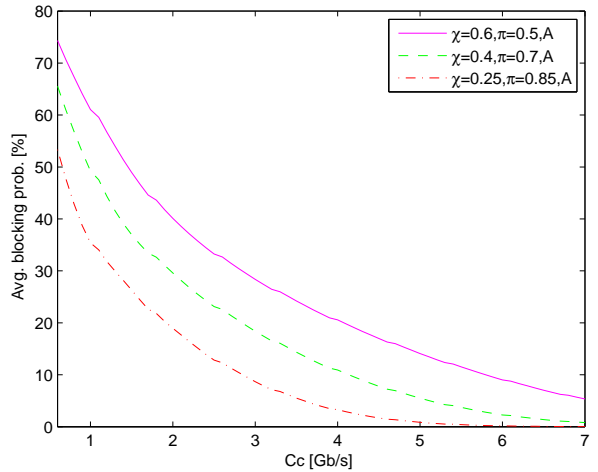}  \\
{\scriptsize (a) Mean request blocking probability $\bar{B}$} \\
\includegraphics[scale=0.615]{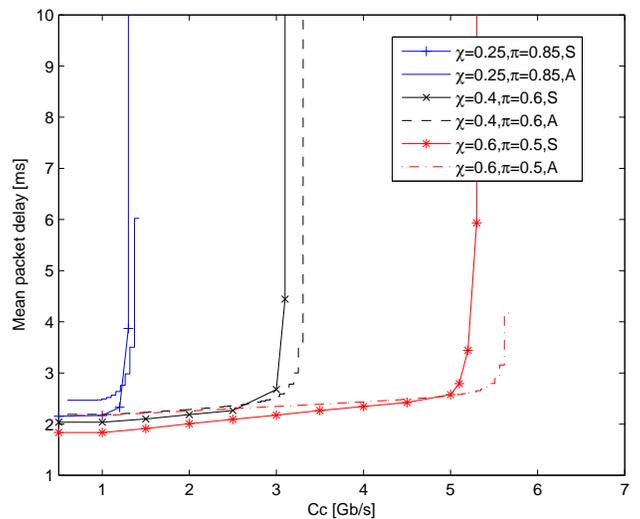}   \\
{\scriptsize (b) Mean packet delay $D$} \\
\end{tabular}
\caption{Impact of circuit service limit $C_c$:
Mean circuit blocking probability $\bar{B}$
(from analysis, Eqn.~(\ref{Bk:eqn}))
and mean packet delay $D$ (from analysis and simulation)
as a function of transmission rate limit for circuit service $C_c$;
fixed mean circuit holding time $1/\mu = 0.02$~s. }
\label{fig:Cc}
\end{figure}
For the packet traffic, we observe from
Table~\ref{fig:chi} a very slight increase in the mean packet delays $D$
as the circuit traffic load $\chi$ increases.
This is mainly because the transmission rate limit $C_c$ for circuit
service bounds the upstream transmission bandwidth the circuits
can occupy to no more than $C_c$ in each cycle.
As the circuit traffic load $\chi$ increases, the circuit traffic
utilizes this transmission rate limit $C_c$ more and more fully.
However, the packet traffic is guaranteed a portion
$1 - C_c/C$ of the upstream transmission bandwidth.
Formally, as the circuit traffic load $\chi$ grows large
($\chi \rightarrow \infty$), the
mean aggregate circuit bandwidth $\bar{\beta}$ approaches the limit
$C_c$, resulting in a lower bound for the packet traffic load
limit~(\ref{pimax:eqn}) of
$\pi_{\max} = 1 - \max\{ 2 \tau/ \Gamma,\ C_c/C \}
- (\omega_o + \omega_u) / \Gamma $
and corresponding upper bounds for the
effective packet traffic intensity $\pi_{\rm eff}$ and
the mean packet delay $D$.

\subsection{Impact of Limit $C_c$ for Circuit Service}
\label{Cc_impact:sec} In Fig.~\ref{fig:Cc} we examine the impact of
the transmission rate limit $C_c$ for circuit traffic. We consider
different compositions $\chi,\ \pi$ of the total traffic load $\chi
+ \pi = 1.05$. We observe from Fig.~\ref{fig:Cc}(a) that the average
circuit blocking probability $\bar{B}$ steadily decreases for
increasing $C_c$. In the example in Fig.~\ref{fig:Cc}, the average
circuit blocking probability $\bar{B}$ drops to negligible values
below 1~\% for $C_c$ values corresponding to roughly twice the
offered circuit traffic load $\chi$. For instance, for circuit load
$\chi = 0.25$, $\bar{B}$ drops to 0.9~\% for $C_c = 5$~Gb/s. The
limit $C_c$ thus provides an effective parameter for controlling
the circuit blocking probability experienced by customers.

From Fig.~\ref{fig:Cc}(b), we observe that the mean packet delay
abruptly increases when the $C_c$ limit reduces the packet traffic
portion $1 - C_c/C$ of the upstream transmission bandwidth to values
near the packet traffic intensity $\pi$. We also observe from
Fig.~\ref{fig:Cc}(b) that the approximate packet delay analysis is
quite accurate for small to moderate $C_c$ values (the slight delay
overestimation is due to neglecting the low packet traffic polling),
but underestimates the packet delays for large $C_c$. Large circuit
traffic limits $C_c$ give the circuit traffic more flexibility for
causing fluctuations of the occupied circuit bandwidth, which
deteriorate the packet service. Summarizing, we see from
Fig.~\ref{fig:Cc}(b) that as the effective packet traffic intensity
$\pi / (1-C_c/C)$ approaches one, the mean packet delay increases
sharply. Thus, for ensuring low-delay packet service, the limit
$C_c$ should be kept sufficiently below $(1-\pi)C$.

When offering circuit and packet service over shared
PON upstream transmission bandwidth, network service providers need to
trade off the circuit blocking probabilities and packet delays.
As we observe from Fig.~\ref{fig:Cc}, the circuit bandwidth limit $C_c$
provides an effective tuning knob for controlling this trade-off.

\subsection{Impact of Low-Packet-Traffic Mode Polling}
\label{Lowtraffic:sec}
\begin{figure}[t]
\includegraphics[scale=0.65]{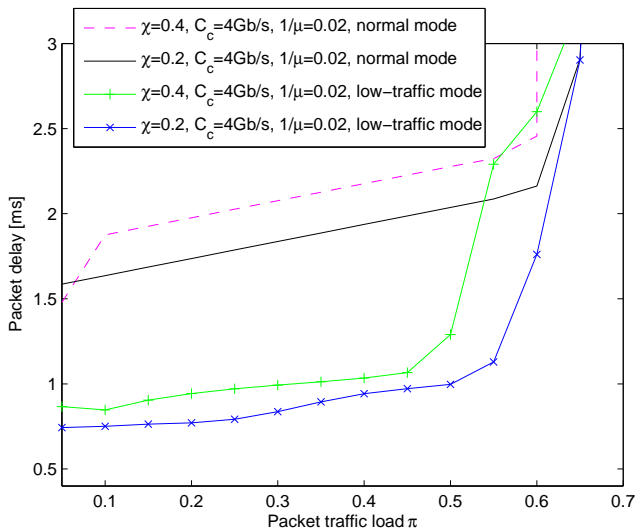}
\caption{Impact of low-packet-traffic polling mode:
Mean packet delay $D$ as a function of packet traffic load $\pi$.
}
\label{fig:LM}
\end{figure}
The Fig.~\ref{fig:LM} we examine the impact of
low-packet-traffic mode polling from Section~\ref{lowload:sec}
on the mean packet delay $D$.
We observe from Fig.~\ref{fig:LM} that low-packet-traffic mode
polling substantially reduces the mean packet delay
compared to conventional polling for low packet traffic loads.
This delay reduction is achieved by the
the more frequent polling which serves packets quicker in cycles with
low load due to circuit traffic.

\section{Conclusion}
\label{sec:conclusion}
We have proposed and evaluated DyCaPPON, a passive optical network
that provides dynamic circuit and packet service.
DyCaPPON is based on fixed duration cycles, ensuring
consistent circuit service, that is completely unaffected by
the packet traffic load.
DyCaPPON masks the round-trip propagation delay for
polling of the packet traffic queues in the ONUs with the
upstream circuit traffic transmissions, providing for efficient
usage of the upstream bandwidth.
We have analyzed the circuit level performance,
including the circuit blocking probability and delay
experienced by circuit traffic in DyCaPPON, as well as
the bandwidth available for packet traffic after serving the
circuit traffic.
We have also conducted an approximate analysis of
the packet level performance.

Through extensive numerical investigations based on
the analytical performance characterization of DyCaPPON as well as
verifying simulations, we have demonstrated the
circuit and packet traffic performance and trade-offs in DyCaPPON.
The provided analytical performance characterizations as well
as the identified performance trade-offs provide tools and guidance
for dimensioning and operating PON access networks that
provide a mix of circuit and packet oriented service.

There are several promising directions for
future research on access networks that flexibly
provide both circuit and packet service.
One important future research direction is to broadly
examine cycle-time structures and wavelength assignments in PONs
providing circuit and packet service.
In particular, the present study focused on a single upstream
wavelength channel operated with a fixed polling cycle duration.
Future research should examine the trade-offs arising from
operating multiple upstream wavelength channels and combinations
of fixed- or variable-duration polling cycles.
An exciting future research
direction is to extend the PON service further toward the
individual user, e.g., by providing circuit and packet service
on integrated PON and wireless access networks,
such as~\cite{AuLMR14,Coim13,DhHJ11,LiKK13,MaGR09,Morad13},
that reach individual mobile users or
wireless sensor networks~\cite{HoH11,SeR11,YuZD12}.
Further, exploring combined circuit and packet service in
long-reach PONs with very long round trip propagation delays,
which may require special protocol mechanisms,
see e.g.,~\cite{Mou05,MeMR13,SKM0110},
is an open research direction.
Another direction is to examine the integration and interoperation
of circuit and packet service in the PON access network with
metropolitan area
networks~\cite{BiBC13,MaRe04,MaRW03,ScMRW03,YaMRC03,YuCL10}
and wide area networks to provide
circuit and packet service~\cite{CircuitSONET}.

\vspace{\baselineskip}

\noindent \textsc{Appendix: Evaluation of Equilibrium Probabilities
$q(\beta)$}

In this Appendix, we present the recursive
Kaufman-Roberts algorithm~\cite[p. 23]{Ross95} for computing the
equilibrium probabilities $q(\beta),\ 0 \leq \beta \leq C_c$ that
the currently active circuit occupy an aggregated bandwidth $\beta$.
For the execution of the algorithm, the given circuit bandwidths
$b_1, b_2, \ldots, b_K$ and limit $C_c$ are suitably normalized so
that incrementing $\beta$ in integer steps covers all possible
combinations of the circuit bandwidth. For instance, in the
evaluation scenario considered in Section~\ref{eval_setup:sec}, all
circuit bandwidth are integer multiples of 52 Mb/s. Thus, we
normalize all bandwidths by 52 Mb/s and for e.g., $C_c = 5$~Gb/s execute
the following algorithm for $\beta = 0, 1, 2, \ldots, 96$. (The
variables $b_k,\ C_c$, and $\beta$ refer to their normalized values,
e.g., $C_c = 96$ for the $C_c = 5$~Gb/s example, in the algorithm below).

The algorithm first evaluates unnormalized occupancy probabilities
$g(\beta)$ that relate to a product-form solution of the
stochastic knapsack~\cite{Ross95}. Subsequently the normalization
term $G$ for the occupancy probabilities is evaluated, allowing then
the evaluation of the actual occupancy probabilities $q(\beta)$.

1. Set $g(0) \gets 1$ and $g(\beta) \gets 0$ for $\beta < 0$.

2. For $\beta=1,2, \ldots, C_c$, set
\begin{eqnarray}
       g(\beta) \gets \frac{1}{\beta} \sum_{k=1}^{K}
      \frac{b_k p_k \lambda_c}{\mu}  g(\beta-b_k).
\end{eqnarray}

3. Set
\begin{eqnarray}
     G = \sum_{\beta = 0}^{C_c} g(\beta).
\end{eqnarray}

4. For $\beta = 0,1, \ldots, C_c$, set
\begin{eqnarray}
      q(\beta) \gets \frac{g(\beta)}{G}.
\end{eqnarray}

\bibliographystyle{IEEEtran}


\end{document}